\documentstyle[12pt]{article}
\def\be{\begin{equation}}
\def\ee{\end{equation}}
\def\bea{\begin{eqnarray}}
\def\eea{\end{eqnarray}}

\begin{document}

\begin{center}
{\Large{\bf Bosonization and Fermionization of the Superstring
Oscillators}}

\vskip .5cm
{\large Davoud Kamani}
\vskip .1cm
{\it Faculty of
 Physics, Amir Kabir University of Technology\\
 P.O.Box: 15875-4413, Tehran, Iran}\\
{\it e-mail: kamani@aut.ac.ir}\\
\end{center}

\begin{abstract}

In this manuscript we consider the transformations of the
oscillators of the bosonic fields of the superstring in terms of
the fermions oscillators and vice versa. We demand the exchange of
the commutation and anti-commutation relations of the
oscillators. Therefore, we obtain some conditions on the
Grassmannian matrices that appear in these transformations. We
observe that there are several methods to obtain these
conditions. In addition, adjoints of the matrix elements and
$T$-duality of these matrices will be obtained. The effects of
this bosonization and fermionization on the mass operators and on
some massless states will be studied. The covariant formalism
will be used and hence we consider both the matter parts and the
ghost parts of the superstring theory.

\end{abstract}
\vskip .5cm



\newpage
\section{Introduction}
There have been some efforts to understand bosonization and
fermionization of various models. In the method of bosonization,
the fermions of a theory are expressed in terms of bosons. This
provides a powerful non-perturbative tool for investigations in
quantum field theory \cite{1}. The extension of this idea can be
seen in the Refs.\cite{2,3,4,5}. Also see \cite{6} and references
therein. The bosonization formalism includes the topics like
smooth bosonization \cite{3}, connection between dual
transformations and bosonization \cite{4}, bosonization in higher
dimensions \cite{5}, functional integral method in bosonization
and so on. In the method of fermionization, one can eliminate the
bosons in favor of fermions. For example, this method has
applications in the heterotic string \cite{7}. In fact, many
phenomena that are difficult to understand in the usual language
take simple forms in the transformed language.

We consider the superstring theory. The transformations of the
worldsheet bosonic fields in terms of the fermionic fields of the
worldsheet (and vice versa) are not consistent. These
inconsistencies appear in the level of the superstring
oscillators. In other words, in this level the transformation
matrices should depend on the mode numbers. Therefore, we
concentrate on the oscillators of the worldsheet fields. Since
calculations and results for the closed and open superstrings are
similar, we only consider the closed superstring theory.

We transform the oscillators of the bosonic fields to the
fermions oscillators and vice versa. This bosonization and
fermionization will be done {\it simultaneously}. These exchanges
will be done for both the matter parts and the ghost parts of the
R-R and NS-NS sectors of the superstring theory. Demanding the
exchange of the commutation and anti-commutation relations of the
oscillators, we obtain some conditions on the transformation
matrices. Therefore, successive transformations ($i.e.$,
rebosonization and refermionization) lead to the trivial or
orthogonal rotations of the oscillators. The elements of the
matrices are Grassmannian variables and carry the mode numbers.
The $T$-duality of these matrices and the adjoints of their
elements will be obtained. The effects of this
bosonization-fermionization on the mass operators and on some
massless states also will be studied.

The exchange of the bosons and fermions oscillators can be
applied on any quantity which is made from the superstring
oscillators. For example, one can apply them on the Virasoro
operators, the BRST charge, boundary state of a D$p$-brane,
string field theory and so on.

This paper is organized as follows. In section 2, we shall study
the effects of our bosonization-fermionization in the matter part
of the R-R sector. In section 3, we shall do the same for the
matter part of the NS-NS sector. In section 4, the $T$-duality of
the transformation matrices will be obtained. In section 5, the
bosonization-fermionization in the ghost parts of the superstring
theory will be discussed. In section 6, the adjoints of the
matrix elements of the transformation matrices will be obtained.
\section{Exchange of oscillators in the matter part of
the R-R sector}
\subsection{Conditions on the transformation matrices}

The quantization of the superstring worldsheet fields leads to the
following equations \bea &~& [\alpha^\mu_m ,
\alpha^\nu_n]=[{\tilde \alpha}^\mu_m , {\tilde
\alpha}^\nu_n]=m\eta^{\mu\nu} \delta_{m+n,0}\;,
\nonumber\\
&~& [\alpha^\mu_m , {\tilde \alpha}^\nu_n]=0\;,
\nonumber\\
&~& [x^\mu , p^\nu]= i \eta^{\mu\nu}\;,
\eea
for the bosonic part, and
\bea
&~& \{d^\mu_m , d^\nu_n\}=\{{\tilde d}^\mu_m , {\tilde d}^\nu_n\}=
\eta^{\mu\nu} \delta_{m+n,0}\;,
\nonumber\\
&~& \{d^\mu_m ,{\tilde d}^\nu_n\}= 0,
\eea
for the fermionic part.

Now we transform the bosons oscillators as in the following \bea
&~& \alpha^\mu_n \longrightarrow
a^\mu_n=(\lambda_n)^\mu_{\;\;\;\nu}d^\nu_n\;,
\nonumber\\
&~& {\tilde \alpha}^\mu_n \longrightarrow {\tilde a}^\mu_n=
({\tilde \lambda}_n)^\mu_{\;\;\;\nu}{\tilde d}^\nu_n\;, \eea
where $n$ is a nonzero integer. For the zero modes we introduce
the transformations \bea &~& lp^\mu \longrightarrow
\lambda^\mu_{\;\;\;\nu}d^\nu_0+ {\tilde
\lambda}^\mu_{\;\;\;\nu}{\tilde d}^\nu_0\;,
\nonumber\\
&~& x^\mu \longrightarrow \chi^\mu_{\;\;\;\nu}d^\nu_0+ {\tilde
\chi}^\mu_{\;\;\;\nu}{\tilde d}^\nu_0\;, \eea where
$l=\sqrt{2\alpha'}$. Assume that the spacetime is non-compact
therefore, $\alpha^\mu_0={\tilde \alpha}^\mu_0=\frac{1}{2}lp^\mu$.
The elements of the matrices $\lambda_n$, ${\tilde \lambda}_n$,
$\lambda$, ${\tilde \lambda}$, $\chi$ and ${\tilde \chi}$ are
Grassmannian variables.

Applying the transformations (3) and (4) in the equations (1) and
using the equations (2), give the following conditions on the
transformation matrices \bea \lambda_n {\lambda}^T_{-n}={\tilde
\lambda}_n {\tilde \lambda}^T_{-n} =-n{\bf 1}\;, \eea for $n \neq
0$. For the zero-mode matrices there are \bea &~& \chi \lambda^T
+ {\tilde \chi}{\tilde \lambda}^T =-il{\bf 1}\;,
\nonumber\\
&~& \lambda \lambda^T + {\tilde \lambda}{\tilde \lambda}^T =0\;,
\nonumber\\
&~& \chi \chi^T + {\tilde \chi}{\tilde \chi}^T =0\;.
\eea
The first condition comes from the noncommutativity of $x^\mu$ and $p^\nu$,
while the second and the third are results of $[p^\mu , p^\nu]=0$ and
$[x^\mu , x^\nu] =0$, respectively.

Now consider the following mappings for the fermions oscillators
in terms of the bosons oscillators \bea &~& d^\mu_n
\longrightarrow
D'^\mu_n=-\frac{1}{n}(\lambda^T_{-n})^\mu_{\;\;\;\nu}
\alpha^\nu_n\;,
\nonumber\\
&~& {\tilde d}^\mu_n \longrightarrow {\tilde D}'^\mu_n =
-\frac{1}{n}({\tilde \lambda}^T_{-n})^\mu_{\;\;\;\nu} {\tilde
\alpha}^\nu_n\;, \eea for $n \neq 0$. The zero modes have the
mappings \bea &~& d^\mu_0 \longrightarrow
\frac{i}{l}\bigg{(}-l(\chi^T)^\mu_{\;\;\;\nu} p^\nu\ +
(\lambda^T)^\mu_{\;\;\;\nu}x^\nu\bigg{)}\;,
\nonumber\\
&~& {\tilde d}^\mu_0 \longrightarrow \frac{i}{l}\bigg{(}
-l({\tilde \chi^T})^\mu_{\;\;\;\nu}p^\nu\ + ({\tilde
\lambda^T})^\mu_{\;\;\;\nu}x^\nu\bigg{)}\;. \eea Since two
successive transformations should be trivial, we require these
forms of transformations for the fermions oscillators. See the
equations (20) and (21). Introducing the bosonizations (7) and (8)
in the equations (2) and using the equations (1) lead to the
conditions \bea \lambda^T_n \lambda_{-n}={\tilde \lambda}^T_n
{\tilde \lambda}_{-n} =n{\bf 1}\;, \eea for the nonzero-mode
matrices, and \bea &~& \chi^T \lambda - \lambda^T\chi =il{\bf
1}\;,
\nonumber\\
&~& \chi^T {\tilde \lambda} - \lambda^T {\tilde \chi} =0\;,
\nonumber\\
&~& {\tilde \chi}^T {\tilde \lambda} - {\tilde \lambda}^T {\tilde
\chi} = il{\bf 1}\;, \eea for the zero-mode matrices. In fact,
the conditions (9) and (10) are not independent of (5) and (6).
The commutators $[\alpha^\mu_0 , \alpha^\nu_n]$ and $[\alpha^\mu_0
, {\tilde \alpha}^\nu_n]$ and the anti-commutators $\{d^\mu_0 ,
d^\nu_n\}$ and $\{d^\mu_0 , {\tilde d}^\nu_n\}$, for $n \neq 0$,
are zero. They do not put new restrictions on the transformation
matrices. We shall see that the conditions (5), (6), (9) and (10)
also can be obtained by various methods.

It is possible to apply only the bosonization or the
fermionization. In this case the transformed theory has one kind
of the commuting or anti-commuting oscillators. However, we
consider both of them. Therefore, on any expression of the
superstring oscillators, the mappings of bosonization and
fermionization will apply {\it simultaneously}.
\subsection{Successive transformations}

We replaced the oscillators $\alpha^\mu_n$, ${\tilde
\alpha}^\mu_n$, $d^\mu_n$ and ${\tilde d}^\mu_n$ with $a^\mu_n$,
${\tilde a}^\mu_n$, $D'^\mu_n$ and ${\tilde D}'^\mu_n$,
respectively. The conditions (5) and (9) imply that the
oscillators $\{a^\mu_n, {\tilde a}^\mu_n, D'^\mu_n, {\tilde
D}'^\mu_n\}$ also obey the equations (1) and (2). Now perform
transformations on the new oscillators \bea &~& a^\mu_n
\longrightarrow A^\mu_n= ({\bar
\lambda}_n)^\mu_{\;\;\;\nu}D'^\nu_n\;,
\nonumber\\
&~& {\tilde a}^\mu_n \longrightarrow {\tilde A}^\mu_n= ({\tilde
{\bar \lambda}}_n)^\mu_{\;\;\;\nu}{\tilde D}'^\nu_n\;, \eea \bea
&~& D'^\mu_n \longrightarrow D^\mu_n=-\frac{1}{n} ({\bar
\lambda}^T_{-n})^\mu_{\;\;\;\nu} a^\nu_n\;,
\nonumber\\
&~& {\tilde D}'^\mu_n \longrightarrow {\tilde D}^\mu_n =
-\frac{1}{n}({\tilde {\bar \lambda}}^T_{-n})^\mu_{\;\;\;\nu}
{\tilde a}^\nu_n\;. \eea The exchange of the commutation and
anti-commutation relations of the transformed oscillators
$\{a^\mu_n, {\tilde a}^\mu_n, D'^\mu_n, {\tilde D}'^\mu_n\}$
under the above transformations leads to the following conditions
for the matrices ${\bar \lambda}_n$ and ${\tilde {\bar
\lambda}}_n$, \bea &~& {\bar \lambda}_n {\bar
\lambda}^T_{-n}={\tilde {\bar \lambda}}_n {\tilde {\bar
\lambda}}^T_{-n}= -n{\bf 1}\;,
\nonumber\\
&~& {\bar \lambda}^T_n {\bar \lambda}_{-n}={\tilde
{\bar \lambda}}^T_n {\tilde {\bar \lambda}}_{-n}= n{\bf 1}\;.
\eea
These conditions are analog of the equations (5) and (9).

Combination of the transformations (3) and (11) gives
\bea &~&
\alpha^\mu_n \longrightarrow A^\mu_n= -\frac{1}{n}({\bar
\lambda}_n \lambda^T_{-n})^\mu_{\;\;\;\nu}\alpha^\nu_n\;,
\nonumber\\
&~& {\tilde \alpha}^\mu_n \longrightarrow {\tilde A}^\mu_n=
-\frac{1}{n}({\tilde {\bar \lambda}}_n {\tilde
\lambda}^T_{-n})^\mu_{\;\;\;\nu}{\tilde \alpha}^\nu_n\;,
\eea
where the definitions of $D'^\mu_n$ and ${\tilde D}'^\mu_n$ from
(7) were used. In fact, these are rebosonization forms of the
oscillators $\alpha^\mu_n$ and ${\tilde \alpha}^\mu_n$. Now
define the matrices $\Lambda_n$ and ${\tilde \Lambda}_n$ as \bea
&~& \Lambda_n=-\frac{1}{n}({\bar \lambda}_n \lambda^T_{-n})\;,
\nonumber\\
&~& {\tilde \Lambda}_n=-\frac{1}{n}({\tilde {\bar \lambda}}_n
{\tilde \lambda}^T_{-n})\;. \eea These matrices satisfy the
identities \bea \Lambda^T_n \Lambda_{-n}=\Lambda_{-n}\Lambda^T_n
={\tilde \Lambda}^T_n {\tilde \Lambda}_{-n} ={\tilde \Lambda}_{-n}
{\tilde \Lambda}^T_n = {\bf 1}\;, \eea which can be interpreted
as orthogonality-like equations.

Using the definitions of $a^\mu_n$ and ${\tilde a}^\mu_n$ from
(3), the equations (7) and (12) lead to the transformations \bea
&~& d^\mu_n \longrightarrow D^\mu_n= -\frac{1}{n}({\bar
\lambda}^T_{-n} \lambda_n)^\mu_{\;\;\;\nu}d^\nu_n\;,
\nonumber\\
&~& {\tilde d}^\mu_n \longrightarrow {\tilde D}^\mu_n=
-\frac{1}{n}({\tilde {\bar \lambda}}^T_{-n} {\tilde
\lambda}_n)^\mu_{\;\;\;\nu}{\tilde d}^\nu_n\;. \eea These are
refermionizations of $d^\mu_n$ and ${\tilde d}^\mu_n$. The
matrices \bea &~& \Gamma_n=-\frac{1}{n}({\bar \lambda}^T_{-n}
\lambda_n)\;,
\nonumber\\
&~& {\tilde \Gamma}_n=-\frac{1}{n}({\tilde {\bar \lambda}}^T_{-n}
{\tilde \lambda}_n)\;, \eea which appeared in the rotations (17),
also satisfy the orthogonality-like equations \bea \Gamma^T_n
\Gamma_{-n}=\Gamma_{-n}\Gamma^T_n ={\tilde \Gamma}^T_n {\tilde
\Gamma}_{-n} ={\tilde \Gamma}_{-n} {\tilde \Gamma}^T_n = {\bf
1}\;. \eea

Similar discussions also hold for the zero modes.
\subsection{Alternatives for the conditions on the matrices}

In the transformations (3) apply the transformations (7). Thus,
we obtain (14) with ${\bar \lambda}_n= \lambda_n$ and ${\tilde
{\bar \lambda}}_n= {\tilde \lambda}_n$, \bea &~& \alpha^\mu_n
\longrightarrow -\frac{1}{n}(\lambda_n
\lambda^T_{-n})^\mu_{\;\;\;\nu}\alpha^\nu_n\;,
\nonumber\\
&~& {\tilde \alpha}^\mu_n \longrightarrow -\frac{1}{n}({\tilde
\lambda}_n {\tilde \lambda}^T_{-n})^\mu_{\;\;\;\nu}{\tilde
\alpha}^\nu_n\;. \eea These rebosonizations should be trivial.
This triviality imposes the conditions (5) on the matrices
$\lambda_n$ and ${\tilde \lambda}_n$. Similarly, in the
transformations (7) introduce the transformations (3). These lead
to the special form of the transformations (17), \bea &~& d^\mu_n
\longrightarrow -\frac{1}{n}(\lambda^T_{-n}
\lambda_n)^\mu_{\;\;\;\nu}d^\nu_n\;,
\nonumber\\
&~& {\tilde d}^\mu_n \longrightarrow -\frac{1}{n}({\tilde
\lambda}^T_{-n} {\tilde \lambda}_n)^\mu_{\;\;\;\nu}{\tilde
d}^\nu_n\;. \eea Triviality of these refermionizations gives the
conditions (9).

The same is true for the zero modes. That is, application of (8)
in (4) leads to the conditions (6). Also application of (4) in (8)
imposes the conditions (10).

{\it The mass operator}

In the R-R sector, the matter part of the mass operator is \bea
\alpha' M^2_R = \sum^\infty_{n=1}\bigg{(}\alpha_{-n}\cdot
\alpha_n+ {\tilde \alpha}_{-n}\cdot {\tilde \alpha}_n
+n(d_{-n}\cdot d_n +{\tilde d}_{-n}\cdot {\tilde
d}_n)\bigg{)}+a_R\;. \eea Under the mappings (3) and (7) this
operator is invariant. In other words, as another alternative,
this invariance leads to the conditions (5) and (9) for the
matrices $\lambda_n$ and ${\tilde \lambda}_n$.

The equations (16) guarantee the invariance of the bosonic part
of this mass operator under the rotations (14). Similarly, the
invariance of the fermionic part, under the rotations (17), can
be seen by the equations (19).
\section{Exchange of oscillators in the matter part of the
NS-NS sector}
\subsection{Conditions on the transformation matrices}

The quantization of the fermionic fields of the worldsheet gives
the following anti-commutation relations \bea &~& \{b^\mu_r ,
b^\nu_s\}=\{{\tilde b}^\mu_r , {\tilde b}^\nu_s\}= \eta^{\mu\nu}
\delta_{r+s,0}\;,
\nonumber\\
&~& \{b^\mu_r , {\tilde b}^\nu_s\}= 0\;. \eea For this sector we
consider the transformations \bea &~& \alpha^\mu_n
\longrightarrow (\theta_n)^\mu_{\;\;\;\nu}b^\nu_{r_n}\;,
\nonumber\\
&~& {\tilde \alpha}^\mu_n \longrightarrow ({\tilde
\theta}_n)^\mu_{\;\;\;\nu}{\tilde b}^\nu_{r_n}\;, \eea where the
elements of the matrices $\theta_n$ and ${\tilde \theta}_n$ are
Grassmannian variables and the half-integer $r_n$ is defined by
\bea r_n = n-\frac{1}{2}{\rm sgn}(n)\;, \;\;\;\; n \neq 0\;. \eea
The function sgn($n$) denotes the sign of $n$. This implies $r_n +
r_{-n}=0$. Substitute the transformations (24) in the equations
(1) and also use (23), we obtain the conditions \bea \theta_n
{\theta}^T_{-n}={\tilde \theta}_n {\tilde \theta}^T_{-n} =-n{\bf
1}\;. \eea

The transformations of the oscillators of the fermions have the
forms \bea &~& b^\mu_{r_n} \longrightarrow
-\frac{1}{n}(\theta^T_{-n})^\mu_{\;\;\;\nu} \alpha^\nu_n\;,
\nonumber\\
&~& {\tilde b}^\mu_{r_n} \longrightarrow -\frac{1}{n}({\tilde
\theta}^T_{-n})^\mu_{\;\;\;\nu} {\tilde \alpha}^\nu_n\;. \eea
Under these mappings the equations (1) and (23) put the following
restrictions on the matrices $\theta_n$ and ${\tilde \theta}_n$,
\bea \theta^T_n {\theta}_{-n}={\tilde \theta}^T_n {\tilde
\theta}_{-n} =n{\bf 1}\;. \eea These are not independent of (26).
Combination of the transformations (24) and (27) also gives the
conditions (26) and (28). The result is analog of the rotations
(20) and (21).

According to the equations (1) and (23) the
bosonization-fermionization in this sector require the equality
$\delta_{r_m + r_n,0}= \delta_{m+n,0}$. Since $m$ and $n$ are
nonzero integers, for the various cases, $i.e.$, $m,n \geq 1$,
$\;m,n \leq -1$ and $m \geq 1$, $\;n \leq -1$ ( or $m \leq -1$,
$\;n \geq 1$) this equality holds.

The zero modes $x^\mu$ and $p^\mu$ can be fermionized in terms of
the oscillators $\{b^\mu_{1/2} , b^\mu_{-1/2}\}$ or $\{{\tilde
b}^\mu_{1/2} , {\tilde b}^\mu_{-1/2}\}$. For example, for the
first set we can write \bea \left( \begin{array}{c}
x\\
lp
\end{array} \right) \longrightarrow
\left( \begin{array}{cc}
\theta_+ & \theta_-\\
\lambda_+ & \lambda_-
\end{array} \right)
\left( \begin{array}{c}
b_{1/2} \\
b_{-1/2}
\end{array}\right)\;,
\eea where $\theta_+$, $\theta_-$, $\lambda_+$ and $\lambda_-$
are $10 \times 10$ matrices with the anti-commuting elements. The
commutation relations of $x^\mu$ and $p^\mu$ put the following
conditions on the matrices \bea &~& \theta_+ \lambda^T_- +
\theta_- \lambda^T_ + = -il{\bf 1},
\nonumber\\
&~& \theta_+ \theta^T_- + \theta_- \theta^T_ + = 0,
\nonumber\\
&~& \lambda_+ \lambda^T_- + \lambda_- \lambda^T_ + = 0.
\eea
These equations correspond to the relations
$[x^\mu , p^\nu]=i\eta^{\mu\nu}$, $[x^\mu , x^\nu]=0$ and
$[p^\mu , p^\nu]=0$, respectively.

The mass operator for this sector is \bea \alpha' M^2_{NS} =
\sum^\infty_{n=1}\bigg{(}\alpha_{-n}\cdot \alpha_n+ {\tilde
\alpha}_{-n}\cdot {\tilde \alpha}_n +r_n(b_{-r_n}\cdot b_{r_n}
+{\tilde b}_{-r_n}\cdot {\tilde b}_{r_n})\bigg{)}+a_{NS}\;. \eea
This operator under the transformations (24) and (27) takes the
form \bea \alpha' {\bar M}^2_{NS} =
\sum^\infty_{n=1}\bigg{(}(1-\frac{1}{2n}) (\alpha_{-n}\cdot
\alpha_n+{\tilde \alpha}_{-n}\cdot {\tilde \alpha}_n)
+n(b_{-r_n}\cdot b_{r_n}+{\tilde b}_{-r_n}\cdot {\tilde b}_{r_n})
\bigg{)}+a_{NS}\;. \eea The mass operator (32) belongs to the
transformed theory. Thus, it acts on the transformed states.

Note that the transformation of the superstring theory gives a
new theory for the superstring. Under this transformation the
algebras of the oscillators remain invariant. This is similar to
the superstring theory and its $T$-dual theory that the mass
operator is common between them.

The successive transformations with the matrices $\theta_n$,
${\tilde \theta_n}$, ${\bar \theta_n}$ and ${\tilde {\bar
\theta_n}}$ produce rotations which are analog of (14) and (17),
\bea &~& \alpha^\mu_n \longrightarrow -\frac{1}{n}({\bar
\theta}_n \theta^T_{-n})^\mu_{\;\;\;\nu}\alpha^\nu_n\;,
\nonumber\\
&~& {\tilde \alpha}^\mu_n \longrightarrow -\frac{1}{n}({\tilde
{\bar \theta}}_n {\tilde \theta}^T_{-n})^\mu_{\;\;\;\nu}{\tilde
\alpha}^\nu_n\;, \eea \bea &~& b^\mu_{r_n} \longrightarrow
-\frac{1}{n}({\bar \theta}^T_{-n} \theta_n)^\mu_{\;\;\;\nu}
b^\nu_{r_n}\;,
\nonumber\\
&~& {\tilde b}^\mu_{r_n} \longrightarrow -\frac{1}{n}({\tilde
{\bar \theta}}^T_{-n} {\tilde \theta}_n)^\mu_{\;\;\;\nu}{\tilde
b}^\nu_{r_n}\;. \eea The matrices in these rotations are
orthogonal-like. Therefore, the bosonic and the fermionic parts
of the mass operator (31), under these rotations, are invariant.
\subsection{Examples}

\subsubsection{A massless state}

Consider the following massless state with the momentum $k^\mu$,
\bea
|\phi^\mu \rangle = b^\mu_{-1/2} | k; Z\rangle\;.
\eea
If the state $| Z\rangle $ is the NS vacuum, the
above state belongs to the open superstring spectrum. Therefore, it describes
a gauge field. If $| Z\rangle $ is the Ramond vacuum, the state (33) is a
massless state of closed superstring.

Under the bosonization, it changes as \bea |\phi^\mu \rangle
\longrightarrow |{\bar \phi}^\mu \rangle \equiv
(\theta^T_1)^\mu_{\;\;\;\nu}\alpha^\nu_{-1} |k; Z\rangle \;. \eea
The state $\alpha^\mu_{-1} |k; Z\rangle $ is a massive vector
field in the bosonic part of the superstring theory. Let
$\varepsilon_\mu$ be polarization vector of the state (35). Thus,
the vector state $\alpha^\mu_{-1} |k; Z\rangle $ has the
polarization \bea \epsilon_\nu =
\varepsilon_\mu(\theta^T_1)^\mu_{\;\;\;\nu}\;. \eea Since the
components of $\varepsilon_\mu$ are commuting numbers, the
components of $\epsilon_\mu$ are anti-commuting numbers. The mass
of the state $|\phi^\mu \rangle$ should be calculated by the
operator (31), while the operator (32) determines the mass of
$|{\bar \phi}^\mu \rangle$, \bea &~& \alpha' M^2_{NS}|\phi^\mu
\rangle = (\frac{1}{2}+a_{NS}) |\phi^\mu \rangle\;,
\nonumber\\
&~& \alpha' {\bar M}^2_{NS}|{\bar \phi}^\mu \rangle =
(\frac{1}{2}+a_{NS}) |{\bar \phi}^\mu \rangle\;. \eea Under the
bosonization-fermionization we have the exchange $|\phi^\mu
\rangle \longleftrightarrow |{\bar \phi}^\mu \rangle$. The
exchange also occurs in the mass operators
$M^2_{NS}\longleftrightarrow {\bar M}^2_{NS}$. Including the ghost
parts, the normal ordering constant is $a_{NS}=-\frac{1}{2}$.

\subsubsection{Presence of a D$p$-brane}

Consider a D$p$-brane that its worldvolume is along the
directions $\{X^\alpha\}$. Let the set $\{X^i\}$ denote the
coordinates which are perpendicular to the worldvolume of the
brane. The massless states \bea b^\alpha_{-1/2} | k \rangle \;,
\;\;\;\;b^i_{-1/2} | k \rangle\;, \eea are gauge field and
collective coordinates of this D$p$-brane. Under the
transformations (27) these states transform as in the following
\bea &~& b^\alpha_{-1/2} | k \rangle \longrightarrow
(\theta^T_1)^\alpha_{\;\;\;\beta} \alpha^\beta_{-1}|k \rangle +
(\theta^T_1)^\alpha_{\;\;\;j} \alpha^j_{-1}|k \rangle\;,
\nonumber\\
&~& b^i_{-1/2} | k \rangle \longrightarrow
(\theta^T_1)^i_{\;\;\;\beta} \alpha^\beta_{-1}|k \rangle +
(\theta^T_1)^i_{\;\;\;j} \alpha^j_{-1}|k \rangle\;. \eea That is,
the transformed versions of the gauge field and the collective
coordinates of the brane have expressions in terms of the massive
vector field and collective coordinates of the bosonic part of
the theory. Note that by the operator (32), the right-hand-sides
of (40) also are massless, as expected.

\subsubsection{The metric and antisymmetric tensor}

The massless state of the NS-NS sector \bea | \phi^{\mu\nu}\;,
\eta \rangle = (b^\mu_{-1/2}{\tilde b}^\nu_{-1/2} +\eta
b^\nu_{-1/2}{\tilde b}^\mu_{-1/2})| k \rangle\;, \eea describes
the metric $G_{\mu\nu}$ (for $\eta = 1$) and the antisymmetric
tensor $B_{\mu\nu}$ (for $\eta = -1$). After bosonization, it
transforms to the state \bea |{\bar \phi}^{\mu\nu}\;,\eta
\rangle=\bigg{(} (\theta^T_1W{\tilde \theta}_1)^{\mu\nu} + \eta
(\theta^T_1W{\tilde \theta}_1)^{\nu\mu}\bigg{)}| k \rangle\;, \eea
where the matrix $W^{\mu\nu}$ has the definition \bea W^{\mu\nu}
= \alpha^\mu_{-1}{\tilde \alpha}^\nu_{-1} \;. \eea Using the mass
operator (32), we observe that the state (42) is massless.
\section{$T$-duality of the transformation matrices}

Let the direction $X^{\mu_0}$ be compact. Therefore, the
oscillators $\{\alpha^{\mu'}_n, d^{\mu'}_n, b^{\mu'}_r, {\tilde
\alpha}^\mu_n, {\tilde d}^\mu_n, {\tilde b}^\mu_r\}$ with $\mu'
\neq \mu_0$, under the $T$-duality do not change, while the
oscillators $\{\alpha^{\mu_0}_n, d^{\mu_0}_n, b^{\mu_0}_r\}$
transform as \bea &~& \alpha^{\mu_0}_n \longrightarrow
-\alpha^{\mu_0}_n\;,
\nonumber\\
&~& d^{\mu_0}_n \longrightarrow -d^{\mu_0}_n\;,
\nonumber\\
&~& b^{\mu_0}_r \longrightarrow -b^{\mu_0}_r\;. \eea On all
oscillators of the superstring, exchanges of the oscillators
accompanied by the $T$-duality transformations can be applied
successively. For example, let study the R-R sector in detail.

On the oscillator $\alpha^{\mu_0}_n$ apply $T$-duality and then
fermionization. These give \bea \alpha^{\mu_0}_n \longrightarrow
-\alpha^{\mu_0}_n \longrightarrow
-(\lambda_n)^{\mu_0}_{\;\;\;\mu_0}d^{\mu_0}_n
-(\lambda_n)^{\mu_0}_{\;\;\;\mu'}d^{\mu'}_n\;. \eea Now perform
fermionization and then $T$-duality \bea \alpha^{\mu_0}_n
\longrightarrow (\lambda_n)^{\mu_0}_{\;\;\;\mu_0}d^{\mu_0}_n
+(\lambda_n)^{\mu_0}_{\;\;\;\mu'}d^{\mu'}_n \longrightarrow
-(\lambda'_n)^{\mu_0}_{\;\;\;\mu_0}d^{\mu_0}_n
+(\lambda'_n)^{\mu_0}_{\;\;\;\mu'}d^{\mu'}_n \;, \eea where the
matrix $\lambda'_n$ is the $T$-dual of $\lambda_n$. We demand that
the final results to be equal. Thus, we obtain \bea
&~&(\lambda'_n)^{\mu_0}_{\;\;\;\mu_0}
=(\lambda_n)^{\mu_0}_{\;\;\;\mu_0}\;,
\nonumber\\
&~& (\lambda'_n)^{\mu_0}_{\;\;\;\mu'}
=-(\lambda_n)^{\mu_0}_{\;\;\;\mu'}\;. \eea Now apply the
processes $T$-duality-fermionization on the oscillator
$\alpha^{\mu'}_n$. These give the other elements of the matrix
$\lambda'_n$, \bea &~& (\lambda'_n)^{\mu'}_{\;\;\;\mu_0}
=-(\lambda_n)^{\mu'}_{\;\;\;\mu_0}\;,
\nonumber\\
&~& (\lambda'_n)^{\mu'}_{\;\;\;\nu'}
=(\lambda_n)^{\mu'}_{\;\;\;\nu'}\;. \eea The equations (47) and
(48) completely determine all elements of the $T$-dual matrix
$\lambda'_n$. Note that if we apply the $T$-duality-bosonization
processes on the fermions oscillators $d^{\mu_0}_n$ and
$d^{\mu'}_n$, we obtain the same result for the matrix
$\lambda'_n$. For the oscillators of the left-moving fields the
corresponding matrices under the $T$-duality do not change,
$i.e.$, \bea ({\tilde \lambda}'_n)^\mu_{\;\;\;\nu} = ({\tilde
\lambda}_n)^\mu_{\;\;\;\nu}\;. \eea

The $T$-dualities of the zero modes
$p^{\mu_0}=p^{\mu_0}_L+p^{\mu_0}_R$ and $x^{\mu_0}=x^{\mu_0}_L +
x^{\mu_0}_R$ are $\frac{1}{\alpha'}L^{\mu_0}=
p^{\mu_0}_L-p^{\mu_0}_R$ and
$x'^{\mu_0}=x^{\mu_0}_L-x^{\mu_0}_R$, respectively. Under the
fermionization we have \bea &~& \frac{l}{\alpha'}L^{\mu_0}
\longrightarrow -\lambda^{\mu_0}_{\;\;\;\nu} d^\nu_0+ {\tilde
\lambda}^{\mu_0}_{\;\;\;\nu}{\tilde d}^\nu_0\;,
\nonumber\\
&~& x'^{\mu_0} \longrightarrow -\chi^{\mu_0}_{\;\;\;\nu}d^\nu_0+
{\tilde \chi}^{\mu_0}_{\;\;\;\nu}{\tilde d}^\nu_0\;. \eea Note
that $p^{\mu'}$ and $x^{\mu'}$ under the $T$-duality are
invariant. Therefore, introducing the processes
$T$-duality-fermionization on $p^{\mu_0}, p^{\mu'}, x^{\mu_0}$
and $x^{\mu'}$, we observe that the $T$-dual versions of the
matrices $\lambda$, ${\tilde \lambda}$, $\chi$ and ${\tilde
\chi}$ also obey the equations (47)-(49).

In the same way, for the NS-NS sector the transformation matrices
under the $T$-duality have the following elements \bea &~&
(\theta'_n)^{\mu_0}_{\;\;\;\mu_0}
=(\theta_n)^{\mu_0}_{\;\;\;\mu_0}\;,
\nonumber\\
&~& (\theta'_n)^{\mu_0}_{\;\;\;\mu'} =-(\theta_n)^{\mu_0}_{\;\;\;\mu'}\;,
\nonumber\\
&~& (\theta'_n)^{\mu'}_{\;\;\;\mu_0} =-(\theta_n)^{\mu'}_{\;\;\;\mu_0}\;,
\nonumber\\
&~& (\theta'_n)^{\mu'}_{\;\;\;\nu'} =(\theta_n)^{\mu'}_{\;\;\;\nu'}\;,
\nonumber\\
&~& ({\tilde \theta}'_n)^\mu_{\;\;\;\nu} = ({\tilde
\theta}_n)^\mu_{\;\;\;\nu}\;. \eea In other words, the
$T$-dualities of the transformation matrices in the R-R and NS-NS
sectors obey the same rule.
\section{Exchange of oscillators in the ghost parts}
\subsection{The ghosts of the R-R sector}
The oscillators of the conformal ghosts and super-conformal
ghosts are $\{b_n , c_n, {\tilde b}_n , {\tilde c}_n \}$ and
$\{\beta_n , \gamma_n, {\tilde \beta}_n , {\tilde \gamma}_n \}$,
respectively. Define the two-component vectors $B^i_n$, ${\tilde
B}^i_n$, $\eta^i_n$ and ${\tilde \eta}^i_n$ with $i \in \{1,2\}$
as in the following
\bea &~& B^1_n = b_n \;,\; B^2_n = c_n\;,
\nonumber\\
&~& {\tilde B}^1_n = {\tilde b}_n \;,\; {\tilde B}^2_n = {\tilde c}_n\;,
\nonumber\\
&~& \eta^1_n = \beta_n \;,\; \eta^2_n = \gamma_n\;,
\nonumber\\
&~& {\tilde \eta}^1_n = {\tilde \beta}_n \;,\;{\tilde \eta}^2_n
={\tilde \gamma}_n\;.
\eea
According to these definitions the commutation and anti-commutation
relations of the ghosts oscillators take
the forms
\bea
&~& [\eta^i_m , \eta^j_n]=[{\tilde \eta}^i_m , {\tilde \eta}^j_n]
=\rho^{ij} \delta_{m+n,0}\;,
\nonumber\\
&~& [\eta^i_m , {\tilde \eta}^j_n] =0\;,
\eea
\bea
&~& \{B^i_m , B^j_n\}=\{{\tilde B}^i_m ,{\tilde B}^j_n\}
= S^{ij}\delta_{m+n,0}\;,
\nonumber\\
&~& \{B^i_m , {\tilde B}^j_n\} =0\;,
\eea
where the symmetric matrix $S$ and the antisymmetric
matrix $\rho$ are defined by
\bea
S= \left( \begin{array}{cc}
0 & 1\\
1 & 0
\end{array} \right)\;,\;\;\;\;
\rho=\left( \begin{array}{cc}
0 & -1\\
1 & 0
\end{array}\right)\;.
\eea
We shall use the following properties of these matrices
\bea
\rho S + S \rho =0\;\;,\;\;\;S^2 = {\bf 1}\;\;,\;\;\;\rho^2=-{\bf 1}\;.
\eea

For the components of the vectors $\eta^i_n$ and ${\tilde
\eta}^i_n$ we introduce the following fermionizations \bea &~&
\eta^i_n \longrightarrow (\Omega_n)^i_{\;\;\;j}B^j_n\;,
\nonumber\\
&~& {\tilde \eta}^i_n \longrightarrow ({\tilde
\Omega}_n)^i_{\;\;\;j} {\tilde B}^j_n\;, \eea where $n$ is any
integer mode number, which includes zero. The elements of the $2
\times 2$ matrices $\Omega_n$ and ${\tilde \Omega}_n$ are
Grassmannian quantities. The equations (53) and (54) and the
fermionization (57) put the following conditions on the matrices
$\Omega_n$ and ${\tilde \Omega}_n$, \bea \Omega_n S
\Omega^T_{-n}={\tilde \Omega}_n S {\tilde \Omega}^T_{-n}=-\rho\;.
\eea

The bosonizations of the components of the vectors $B^i_n$ and
${\tilde B}^i_n$ can be considered as \bea &~& B^i_n
\longrightarrow (S \Omega^T_{-n}\rho)^i_{\;\;\;j}\eta^j_n\;,
\nonumber\\
&~& {\tilde B}^i_n \longrightarrow (S {\tilde
\Omega}^T_{-n}\rho)^i_{\;\;\;j} {\tilde \eta}^j_n\;. \eea
Introducing these mappings into (54) and using (53) lead to the
conditions \bea \Omega^T_n \rho \Omega_{-n}={\tilde \Omega}^T_n
\rho {\tilde \Omega}_{-n}=S\;. \eea These conditions are not
independent of (58). That is, combinations of the equations (58)
and (60) are trivial identities.

The contributions of the super-conformal ghosts and conformal
ghosts to the mass operator are \bea \alpha' M^2_{scg} = -
\sum^\infty_{n=1}\bigg{(} n(\eta^T_{-n}\rho \eta_n + {\tilde
\eta}^T_{-n}\rho {\tilde \eta}_n)\bigg{)}\;, \eea \bea \alpha'
M^2_{cg} =  \sum^\infty_{n=1}\bigg{(} n(B^T_{-n} S B_n + {\tilde
B}^T_{-n} S {\tilde B}_n)\bigg{)}\;. \eea Now introduce the
transformations (57) and (59) into these operators. According to
the conditions (58) and (60) the masses $\alpha' M^2_{cg}$ and
$\alpha'M^2_{scg}$ are exchanged. In other words, the quantity
$\alpha'( M^2_{cg}+ M^2_{scg})$ remains invariant.
\subsection{Successive transformations}

Combine the transformations (57) and (59). We obtain analog of the
equations (20) and (21), \bea &~& \eta^i_n \longrightarrow
(\Omega_n S \Omega^T_{-n}\rho)^i_{\;\;\;j} \eta^j_n\;,
\nonumber\\
&~&  B^i_n \longrightarrow (S \Omega^T_{-n}\rho
\Omega_n)^i_{\;\;\;j} B^j_n\;, \eea which are rebosonization and
refermionization. Similar transformations also hold for the
oscillators of the left-moving fields. Triviality of these
transformations leads to the conditions (58) and (60). Inversely,
the conditions (58) and (60) imply that the above mappings to be
trivial.

Now consider two different successive transformations. The first
is distinguished by the matrices $\Omega_n$ and ${\tilde
\Omega}_n$, and the second by ${\bar \Omega}_n$ and ${\tilde
{\bar \Omega}}_n$. The matrices ${\bar \Omega}_n$ and ${\tilde
{\bar \Omega}}_n$ also satisfy the conditions (58) and (60).
Therefore, we have non-trivial rebosonization and refermionization
\bea &~& \eta^i_n \longrightarrow ({\bar \Omega}_n S
\Omega^T_{-n}\rho)^i_{\;\;\;j} \eta^j_n \equiv
(M_n)^i_{\;\;\;j}\eta^j_n\;,
\nonumber\\
&~&  B^i_n \longrightarrow (S {\bar \Omega}^T_{-n}\rho
\Omega_n)^i_{\;\;\;j} B^j_n \equiv (N_n)^i_{\;\;\;j}B^j_n\;. \eea
These are analog of the transformations (14), (17), (33) and (34).
Similar relations also hold for the left-oscillators. The
matrices $M_n$ and $N_n$ satisfy the identities \bea M^T_n \rho
M_{-n}=M_n \rho M^T_{-n}= \rho \;, \eea \bea N^T_n S N_{-n}=N_n S
N^T_{-n}= S\;. \eea The meaning of the equations (65) is
invariance of the mass (61) under the first transformation of
(64). Also the equations (66) imply that the mass (62) under the
second transformation of (64) is invariant.
\subsection{The ghosts of the NS-NS sector}
For the super-conformal ghosts of this sector we have the vectors
$\eta^i_r$ and ${\tilde \eta}^i_r$, with the components \bea &~&
\eta^1_r = \beta_r\;,\;\;\;\eta^2_r=\gamma_r\;,
\nonumber\\
&~& {\tilde \eta}^1_r = {\tilde \beta}_r\;,\;\;\;{\tilde \eta}^2_r
={\tilde \gamma}_r\;, \eea where $r$ is a half-integer number. The
commutation relations of these oscillators are \bea &~& [\eta^i_r
, \eta^j_s]=[{\tilde \eta}^i_r , {\tilde \eta}^j_s] =\rho^{ij}
\delta_{r+s,0}\;,
\nonumber\\
&~& [\eta^i_r , {\tilde \eta}^j_s] =0\;.
\eea

Let the transformation matrices for this sector be $\omega_n$ and
${\tilde \omega}_n$. Therefore, we have the following
fermionizations \bea &~& \eta^i_{r_n} \longrightarrow
(\omega_n)^i_{\;\;\;j}B^j_n\;,
\nonumber\\
&~& {\tilde \eta}^i_{r_n} \longrightarrow ({\tilde
\omega}_n)^i_{\;\;\;j} {\tilde B}^j_n\;, \eea where the
half-integer $r_n$ has been given by the definition (25). The
bosonizations of $B^i_n$ and ${\tilde B}^i_n$, for $n \neq 0$,
are \bea &~& B^i_n \longrightarrow (S
\omega^T_{-n}\;\rho)^i_{\;\;\;j}\eta^j_{r_n}\;,
\nonumber\\
&~& {\tilde B}^i_n \longrightarrow (S {\tilde
\omega}^T_{-n}\;\rho)^i_{\;\;\;j} {\tilde \eta}^j_{r_n}\;. \eea
According to the mappings (69) and (70), the relations (54) and
(68) (or triviality of the successive transformations analog of
(63)) give the following conditions on the matrices $\omega_n$
and ${\tilde \omega}_n$, \bea &~& \omega_n S \omega^T_{-n}={\tilde
\omega}_n S {\tilde \omega}^T_{-n} =-\rho\;,
\nonumber\\
&~&\omega^T_n \rho \omega_{-n}={\tilde \omega}^T_n \rho {\tilde
\omega}_{-n}=S\;. \eea These conditions are not independent of
each other. Since the features of the commutation relations (53)
and (68) are the same, the matrices $\omega_n$ and $\Omega_n$
obey the same conditions.

For the zero modes we consider the transformations
\bea
&~& \left( \begin{array}{c}
B^1_0\\
B^2_0
\end{array} \right) \longrightarrow
S\omega^T_0 \rho
\left( \begin{array}{c}
\eta^1_{1/2} \\
\eta^2_{-1/2}
\end{array}\right)\;,
\nonumber\\
&~& \left( \begin{array}{c}
{\tilde B}^1_0\\
{\tilde B}^2_0
\end{array} \right) \longrightarrow
S{\tilde \omega}^T_0 \rho
\left( \begin{array}{c}
{\tilde \eta}^1_{1/2} \\
{\tilde \eta}^2_{-1/2}
\end{array}\right)\;.
\eea The commutation relations (54) for $m=n=0$ give the following
conditions on the matrices $\omega_0$ and ${\tilde \omega}_0$,
\bea \omega^T_0 \rho \omega_0= {\tilde \omega}^T_0 \rho {\tilde
\omega}_0= S. \eea

The conformal ghost part of the mass operator has been given by
(62). The contribution of the super-conformal ghosts is \bea
\alpha' M^2_{scg} = - \sum^\infty_{n=1}\bigg{(} r_n(\eta^T_{-r_n}
\rho \eta_{r_n} + {\tilde \eta}^T_{-r_n}\rho {\tilde
\eta}_{r_n})\bigg{)}\;. \eea  The sum of the masses (62) and (74)
under the transformations (69) and (70) changes to \bea \alpha'
({\bar M}^2_{cg}+{\bar M}^2_{scg}) = \sum^\infty_{n=1}\bigg{(}
-n(\eta^T_{-r_n}\rho \eta_{r_n} + {\tilde \eta}^T_{-r_n}\rho
{\tilde \eta}_{r_n}) + r_n(B^T_{-n}S B_n + {\tilde B}^T_{-n}S
{\tilde B}_n)\bigg{)}\;. \eea This operator acts on the
transformed states, under (69) and (70).

There are equations analog of (64)-(66) with $\omega_n$ and
${\bar \omega}_n$ instead of $\Omega_n$ and ${\bar \Omega}_n$. By
these equations, the operators $\alpha' M^2_{cg}$ and $\alpha'
M^2_{scg}$ are invariant.
\section{Adjoint of a matrix element}

The adjoint of a superstring oscillator has the property \bea
(q^\mu_k)^\dagger =q^\mu_{-k}\;, \eea where $q^\mu_k$ denotes any
oscillator of the superstring. For the R-R sector $k$ is integer
and for the NS-NS sector it is half-integer. According to this
equation, the matrix elements of the transformation matrices have
the adjoint as in the following \bea &~&
[(Q_n)^\mu_{\;\;\;\nu}]^\dagger = (Q_{-n})^\mu_{\;\;\;\nu}\;,
\nonumber\\
&~& Q_n \in \{\lambda_n , {\tilde \lambda}_n , \theta_n , {\tilde
\theta}_n, \Omega_n , {\tilde \Omega}_n, \omega_n , {\tilde
\omega}_n\}\;. \eea This is due to the uniqueness of the
bosonization and fermionization of the superstring oscillators.
Note that this is adjoint of the matrix elements. It is not the
adjoint of the matrices.
\section{Conclusions and summary}

By using the matrices with the Grassmannian elements, we
considered transformations of the oscillators of the bosonic
fields in terms of the oscillators of the fermionic fields of the
superstring, and vice versa. The exchange of the commutation and
anti-commutation relations of these oscillators (under the above
transformations) puts some conditions on the above matrices. These
conditions also can be obtained from the other methods such as
triviality of the similar successive transformations.

{\it The matter parts}

In the R-R sector the mass operator remains invariant, while in
the NS-NS sector it changes. The transformed mass operator acts
on the transformed states of the superstring. On the other hand,
the transformed theory is a new theory which has common oscillator
algebra with the initial theory.

Different successive transformations give (orthogonal) rotations
to the oscillators. In other words, these rotations are
rebosonization and refermionization. Therefore, under these
rotations the bosonic and fermionic parts of the mass operators of
the both sectors separately remain invariant.

We obtained the polarization of a massive vector field in terms
of the polarization of the superstring gauge field. This vector
field comes from the bosonic part of the superstring theory. In
addition, transformations of the gauge field and the collective
coordinates of a D$p$-brane were analyzed. The new versions of
the spacetime metric and the antisymmetric tensor also were
obtained.

The effects of the $T$-duality on the transformation matrices were
studied. We observed that in the R-R and NS-NS sectors the
$T$-dualities of the Grassmannian matrices obey the same rule.

{\it The ghost parts}

Since we considered the covariant formulation of the superstring
theory, the conformal and the super-conformal ghosts are also
analyzed. Therefore, in each sector, by using the corresponding
bosonization-fermionization, we obtained the conditions on the
transformation matrices and the changes of the mass operators.

We observed that the mass operator in the R-R sector is
invariant, while in the NS-NS sector it changes. The matrices of
the non-trivial rebosonization and refermionization satisfy some
identities. Under these transformations, the mass operators of
the conformal ghosts and super-conformal ghosts separately are
invariant.

Finally, we saw that the adjoint of a matrix element, similar to
the adjoint of an oscillator, changes the sign of the mode number
of that matrix element.

\end{document}